\documentclass[conference]{IEEEtran}
\IEEEoverridecommandlockouts
\usepackage{cite}
\usepackage{amsmath,amssymb,amsfonts}
\usepackage{algorithmic}
\usepackage{graphicx}
\usepackage{textcomp}
\usepackage{xcolor}
\usepackage{soul} 
\sethlcolor{yellow}

\usepackage{marginnote}
\usepackage{todonotes}

\usepackage[normalem]{ulem} 

\def\BibTeX{{\rm B\kern-.05em{\sc i\kern-.025em b}\kern-.08em
    T\kern-.1667em\lower.7ex\hbox{E}\kern-.125emX}}
\begin{document}

\title{Advances in the Fabrication of On-chip Superconducting Integral Field Units for CMB and Line-Intensity Astronomy\\

}
\author{
\IEEEauthorblockN{
    L.G.G. Olde Scholtenhuis$^{1}$,
    D. Perez Capelo$^{2}$,
    K. Karatsu$^{2}$,
    D.J. Thoen$^{2}$,
    A.J. van der Linden$^{2}$,
}
\IEEEauthorblockN{
    S.O. Dabironezare$^{1,2}$,
    L.H. Marting$^{1}$,
    J.J.A. Baselmans$^{1,2,3}$,
    S. Vollebregt$^{1}$, and
    A. Endo$^{1}$
}

\IEEEauthorblockA{
    $^{1}$\textit{Department of Microelectronics, Delft University of Technology}, Delft, the Netherlands\\
    $^{2}$\textit{SRON Space Research Organization Netherlands}, Leiden, the Netherlands\\
    $^{3}$\textit{Physikalisches Institut, Universität zu Köln}, Köln, Germany
}
}

\maketitle

\begin{figure*}[htbp]
    \centering
    \includegraphics[width=0.85\linewidth]{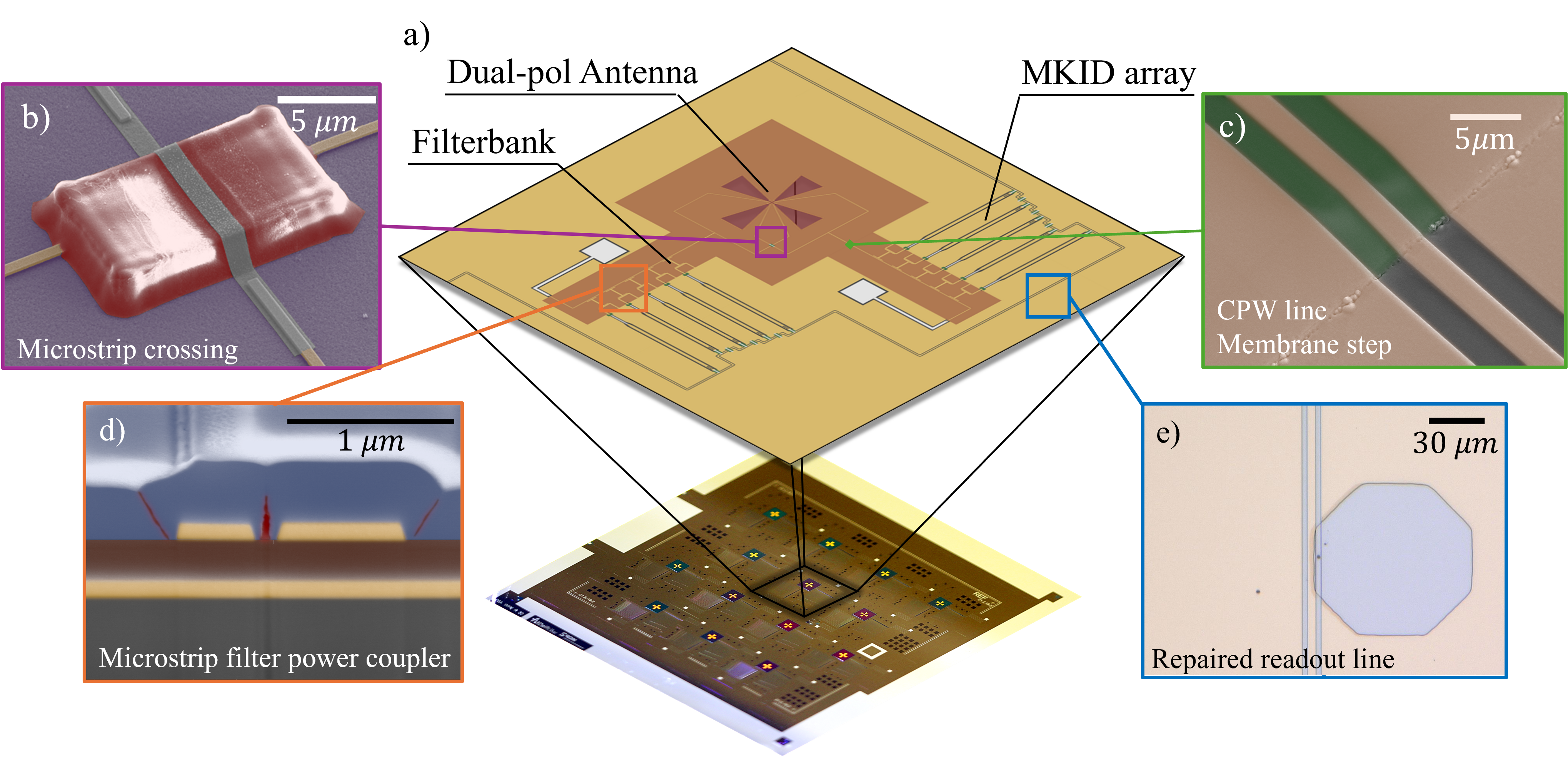}
    \caption{An overview image of the on-chip improvements discussed in this article. \textbf{a.} shows a schematic of a spaxel above a successfully fabricated fourteen-spaxel IFU. The four advances presented in this paper are \textbf{b.} the microstrip crossing to enable dual-pol reception, \textbf{c.} the CPW me mbrane-substrate transition (a. shows a microstripline instead),\textbf{d.} a dielectric-covered microstrip power coupler, and \textbf{e.} a repaired section of readout line. The colored lines indicate their usual on-spaxel position. More detailed images will be presented in the following sections.
    }
    \label{fig:figure_1}
\end{figure*}

\begin{abstract}
Studying the polarization and spectral distortion of the Cosmic Microwave Background (CMB) in tandem with intensity fluctuations of the Cosmic Infrared Background (CIB) allows us to verify our assumptions on cosmic inflation and investigate the dynamics and evolution of galaxy clusters in the last 10 billion years.
Because of its broadband emission and being an all-sky extended source, observing the entire CMB in detail is a very time-consuming and expensive exercise.
Fortunately, in the last few years, the on-chip superconducting spectrometer technology has moved out of the lab and into the telescope. 
With its compact size and background-limited sensitivity, this family of instruments is particularly well-suited for fast and large area observations in a relatively unexplored range of the electromagnetic spectrum.
However, recent examples of this technology do not yet reach the requirements needed for large spectroscopic and polarimetric surveys of the CMB.
We formulate several of these requirements and introduce novel on-chip components and fabrication techniques.
We introduce a cross-over to enable distinguishing signal polarization, minimize signal loss by locally optimized lithography of a coplanar-waveguide (CPW), lower the spectral resolution of microstrip filters by deposition of a dielectric layer, and increase the yield of the spectrometer array by removing individual line shorts. These together have culminated in the successful fabrication of a fourteen-spaxel IFU.
  
\end{abstract}


\section{Introduction}
Sensitive and large spectroscopic and polarimetric surveys of the millimeter-wave sky will be groundbreaking to many areas of cosmology and astrophysics, such as the inflation model of cosmology, the growth of the cosmic large-scale structure, and the formation of stars and galaxies\cite{delabrouilleMicrowaveSpectropolarimetryMatter2021}. For example, precise line-intensity mapping (LIM) measurements can rule out certain models of cosmic inflation \cite{karkareSnowmass2021Cosmic2022}, whereas the growth dynamics of the cosmic large-scale structure and galaxy clusters can be studied in detail by observing the Sunyaev–Zeldovich effect imprinted on the CMB\cite{dimascoloFormingIntraclusterGas2023}. Moreover, these maps will also contain the collective emission of dust and atomic/molecular emission lines from galaxies across the entire cosmic history\cite{kovetzLineIntensityMapping20172017}. Such an observation calls for an instrument with spectral pixels (spaxels) that combine ultra-wide bandwidth (collectively multiple octaves), low spectral resolution ($R = F/dF \approx20$), and polarization sensitivity. Moreover, the instrument should multiplex many of such spaxels for a fast mapping speed.

An integral field unit (IFU) based on the integrated superconducting spectrometer (ISS) technology is a promising architecture for such an instrument\cite{jovanovic2023AstrophotonicsRoadmap2023a}. Unlike conventional ultra-wideband mm-wave spectrometers based on dispersive quasi-optical spectrometers, the ISS integrates both the spectrometer and the detectors onto a single wafer as a monolithic superconducting integrated circuit. A single ISS spaxel is typically a combination of a lens-antenna, a superconducting filterbank for dispersion, and an array of microwave Kinetic Inductance Detectors (MKIDs) that measures the power at the output of each channel of the filterbank. Such a spaxel has a footprint in the order of a $\mathrm{cm}^2$, and can be tiled into a 2D array to form a densely-packed IFU. For example, the DESHIMA 2.0 spectrometer on the ASTE telescope has demonstrated an ISS spaxel that instantaneously covers a full octave of 200-400 GHz \cite{endoFirstLightDemonstration2019}\cite{taniguchiDESHIMA20Development2022}\cite{moermanAlignmentOpticalVerification2025a}. However, DESHIMA 2.0 has only one spaxel, is only sensitive to a single linear polarization, and the spectral resolution of $R = F/dF = 500$ could be lower for a mission targeted towards the CMB. Despite the succesfull fabrication of a single ISSs, the nano/microfabrication for upscaling the ISS technology to an IFU has yet to be demonstrated \cite{hahnleSuperconductingIntegratedCircuits2021}\cite{pascuallagunaOnChipSolutionsFuture2022}.

Here, we present four key advances in the nano/micro-fabrication technology to enable future superconducting IFUs capable of CMB observations: 
\\1) microstrip cross-overs that enable dual-polarization antenna reception, 
\\2) membrane-substrate crossings of a coplanar-waveguide that enable ultra-wideband antenna reception,
\\3) microstrip power couplers with gaps filled with a deposited dielectric that enable low-resolution spectroscopy, and 
\\4) a technique to repair defects in long transmission lines to enable multiplexed readout of many spaxels. 
\\Collectively, these advances have led to a successful fabrication of a fourteen-spaxel, dual-polarization IFU as presented in figure \ref{fig:figure_1} (Karatsu et al., in prep).

\section{Microstrip cross-overs for dual-polarization antenna reception}
To study the polarization, of for example the CMB, IFUs need a way to separate the incoming radiation by its polarization. Here, we look into the use of the dual-polarized Leaky Lens antenna (Dabironezare et al., in prep). This antenna consists of two perpendicularly oriented bow-tie-shaped slot antennas. The two opposite sides of the antenna are fed with microstrip lines that join together before continuing to the filter bank. This, however, requires the two signal lines from the opposing polarizations to cross at some point. This creates a potential source of cross-coupling between the two lines. 

We present a design that allows these two lines to cross. And because the bridges along the CPW readout line have a similar architecture, we can do this without increasing the number of fabrication steps\cite{pascuallagunaOnChipSolutionsFuture2022}.
 
First, a 1.5~$\mathrm{\mu m}$ thick mesa was created by spinning a layer of polyimide on the substrate. Using UV-lithography, we expose a small area around the crossing point where one of the lines is interrupted.
Unlike a dielectric deposited by PECVD, the polyimide support can be patterned without plasma etching and therefore avoids the risk of etching exposed surfaces.
A 40~$\mathrm{nm}$ thick sputtered aluminum line is then patterned on top of the mesa to connect the two sides of the interrupted line. Due to the placement of the aluminum on top of the polyimide and the membrane, this structure slightly protrudes from the rest of the wafer. To prevent the mask from exerting concentrated forces on these structures, which could damage the resist and the aluminum, we used proximity exposure during optical patterning of the aluminum.
We choose aluminum because it is compatible with the existing layer stack, and its wet etching process does not affect the NbTiN lines.
To match the impedance of the aluminum strip to that of the NbTiN line, the strip was further narrowed through additional electron beam exposure and wet etching.  The structures are similar to the DESHIMA CPW-bridge designs that have shown excellent mechanical and thermal stability after repeated cooldowns \cite{pascuallagunaOnChipSolutionsFuture2022}. 

\begin{figure}[htbp]
    \centering
    \includegraphics[width=\linewidth]{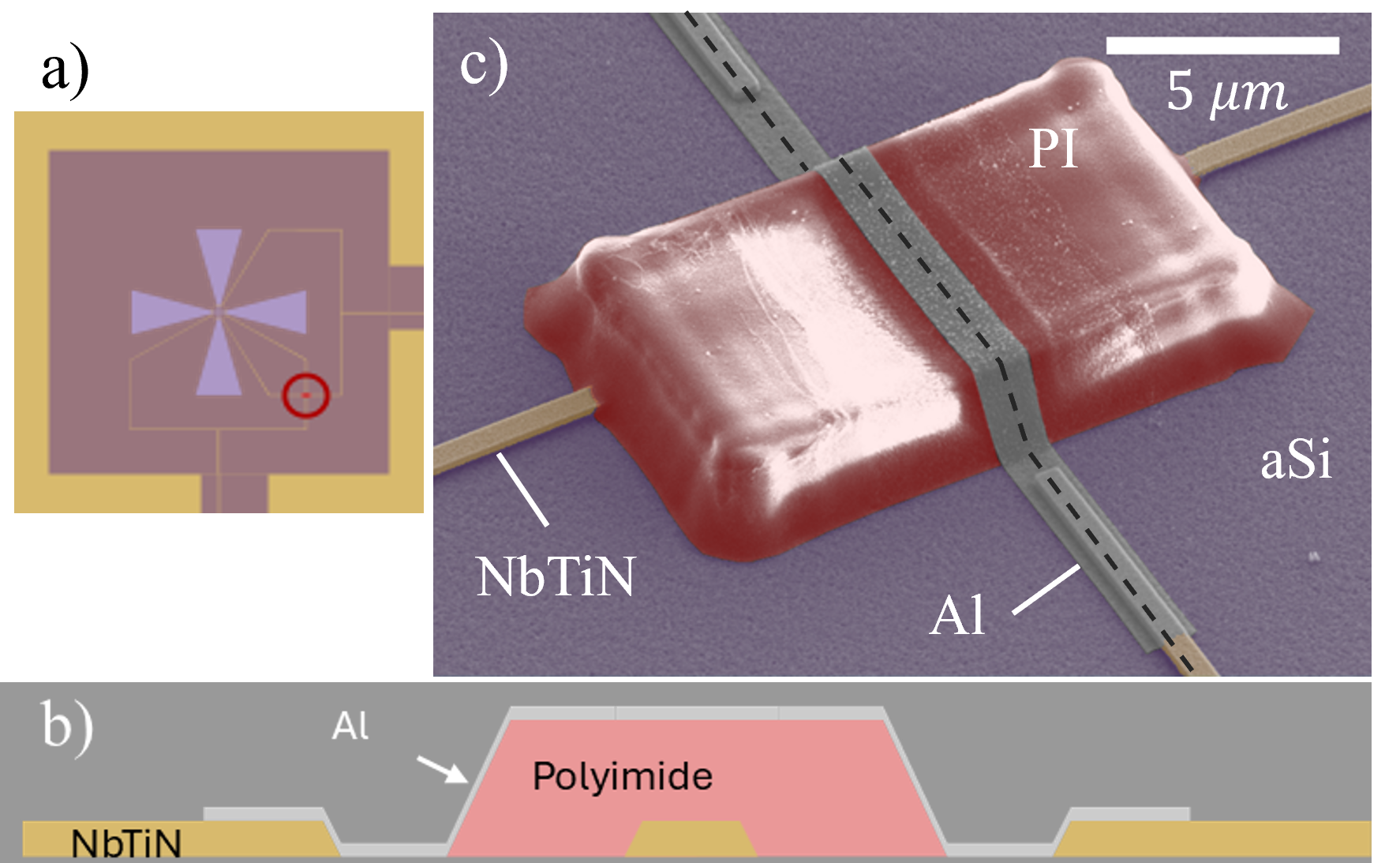}
    \caption{\textbf{a.} The signal lines from both sides of the dual-pol Leaky Lens antenna will inevitably cross, indicated by the circle. \textbf{b.} Using a polyimide (red) and aluminum (light grey) bridge style structure, the lines can cross with minimal cross-coupling. \textbf{c.} SEM image of a coupler that showed excellent transmission. The dashed line indicates the cut of the cross-section in b.
    }
    \label{fig:linecrossing}
\end{figure}

This design was successfully implemented on a spectrometer and showed excellent transmission for both polarizations (Dabironezare et al., in prep). Figure \ref{fig:linecrossing} shows a diagram of the dual-pol-bowtie-antenna with the location of the structure, together with a cross-section of the pattern and a Scanning Electron Microscope (SEM) image of a finished structure.

\section{Cpw membrane-substrate transition for ultrawide instantaneous bandwidth}
Besides its ability to distinguish polarization, a detector array with a large instantaneous bandwidth would significantly speed up observing broadband sky signals.
To do so, an IFU would need an antenna to couple the incoming radiation to the on-chip components.
Although there are several methods to do this, we will focus on the use of a Leaky Lens antenna. Because of its excellent coupling efficiency across an octave bandwidth, this architecture is well-suited for use in an IFU \cite{hahnleUltrawidebandLeakyLens2020}.

This choice, however, implies the fabrication of a thin membrane on top of which a planar antenna is patterned. The fabrication of this membrane is described in more detail in \cite{hahnleSuperconductingIntegratedCircuits2021}.
The relatively lossy SiN of the membrane could lead to significant losses if the distance between the antenna and the filterbank is large. Instead, a CPW-line on the crystalline silicon substrate is preferred. To minimize total losses, the SiN is localized only around the position of the antenna. This, however, introduces a transition in the signal line from the membrane to the substrate, which will introduce additional complexities in the fabrication, especially when using electron beam lithography.

\begin{figure}[htbp]
    \centering
    \includegraphics[width=\linewidth]{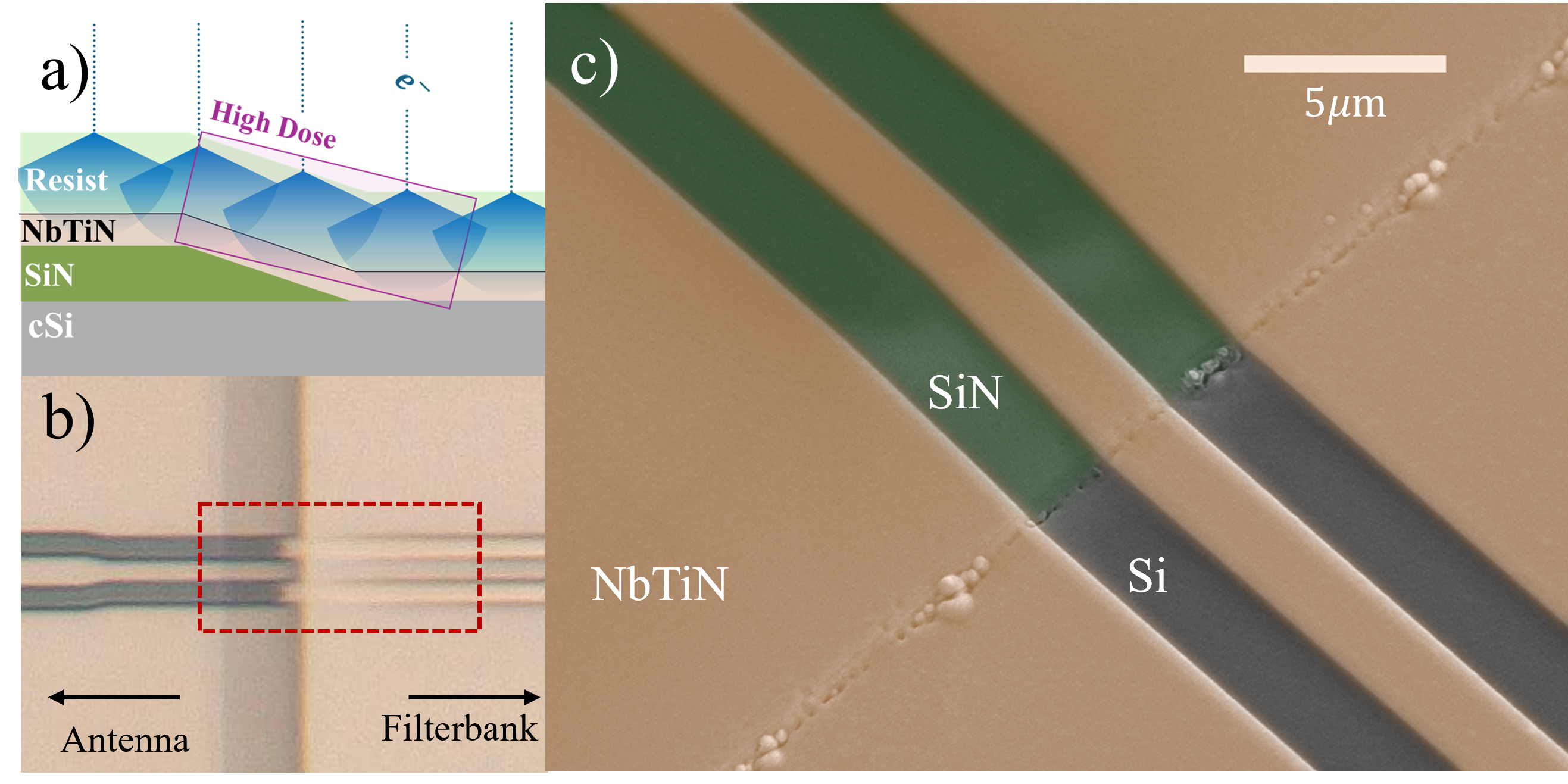}
    \caption{\textbf{a.} The slanted resist layer causes scattered electrons (blue), during e-beam patterning, to overexpose the area around the transition (purple). \textbf{b.} shows an optical image of the step as a result of this overexposure. The red square highlights the overexposed area and the shorted line. \textbf{c.} shows a SEM image of the NbTiN CPW transmission line (orange) across the step at the edge of the SiN membrane (green) after the exposure dose is reduced from 1100 to 500~$\mathrm{\mu C}/\mathrm{cm}^2$ during patterning.
    }
    \label{fig:membrane_step}
\end{figure}

On the sloped section where the CPW descends from the SiN layer down to the Si substrate, the effective electron-beam exposure of the ma-N1405 resist increases due to the sloped surface illustrated in figure \ref{fig:membrane_step}a. As a result, this leads to shorts between the CPW strip and the ground plane as shown in figure \ref{fig:membrane_step}b.
To reach the precision needed for these patterns, we use the method as described in \cite{thoenCombinedUltravioletElectronbeam2022a}. 
An overexposure of the CPW pattern, in the negative tone resist ma-N1405, creates line shorts, which would prevent the signal from traveling towards the filter. Figure \ref{fig:membrane_step}b shows such an example.
Standard compensation techniques, such as a Proximity Effect Correction (PEC), assume continuous layers, disregarding any height variations. Instead, after several dose tests on similar structures, we locally reduced the writing dose by ca. 45\% in the area around the pattern from 1100 to 500~$\mathrm{\mu C}/\mathrm{cm}^2$. This way, the received dose on the step's side is lowered. To minimize the effect of misalignment, we make sure the low-dose pattern has a 0.5~$\mathrm{\mu m}$ overlap with the high dose, away from the step. Figure \ref{fig:membrane_step}c shows an SEM image of the step without shorts.

\section{Low spectral resolution filters for continuum spectroscopy}
For CMB observations, a low spectral resolution of $R$ = 10-20 can be desired to maximize the product of spectral bandwidth and field-of-view for a given detector count. By lowering the $R$ of the microstrip filters, we can keep the same overall instantaneous bandwidth while decreasing the number of detectors. One way to achieve this is by decreasing the quality factor ($Q$) of these microstrip filters and their couplers. 

Because the coupling Quality factor ($Q_c$) is inversely proportional to the capacitance, we would need to increase the capacitance. 
This can be achieved by placing the coupler line closer to the signal line or by increasing the effective dielectric constant of its environment. In the current design, the gap width of the coupler is 250~$\mathrm{nm}$, close to the limit of what we can achieve with our lithography and etching methods without significantly lowering the precision. For that reason, we instead opt for increasing the effective dielectric constant ($\epsilon_{eff}$) by depositing a layer of dielectric material on top of the filter and coupler structures. We deposited an 800~$\mathrm{nm}$ layer of a-Si:H at 250 degrees using an Oxford Plasmalab80+ PECVD for this purpose \cite{buijtendorpCharacterizationLowlossHydrogenated2022}.

\begin{figure}
    \centering
\includegraphics[width=\linewidth]{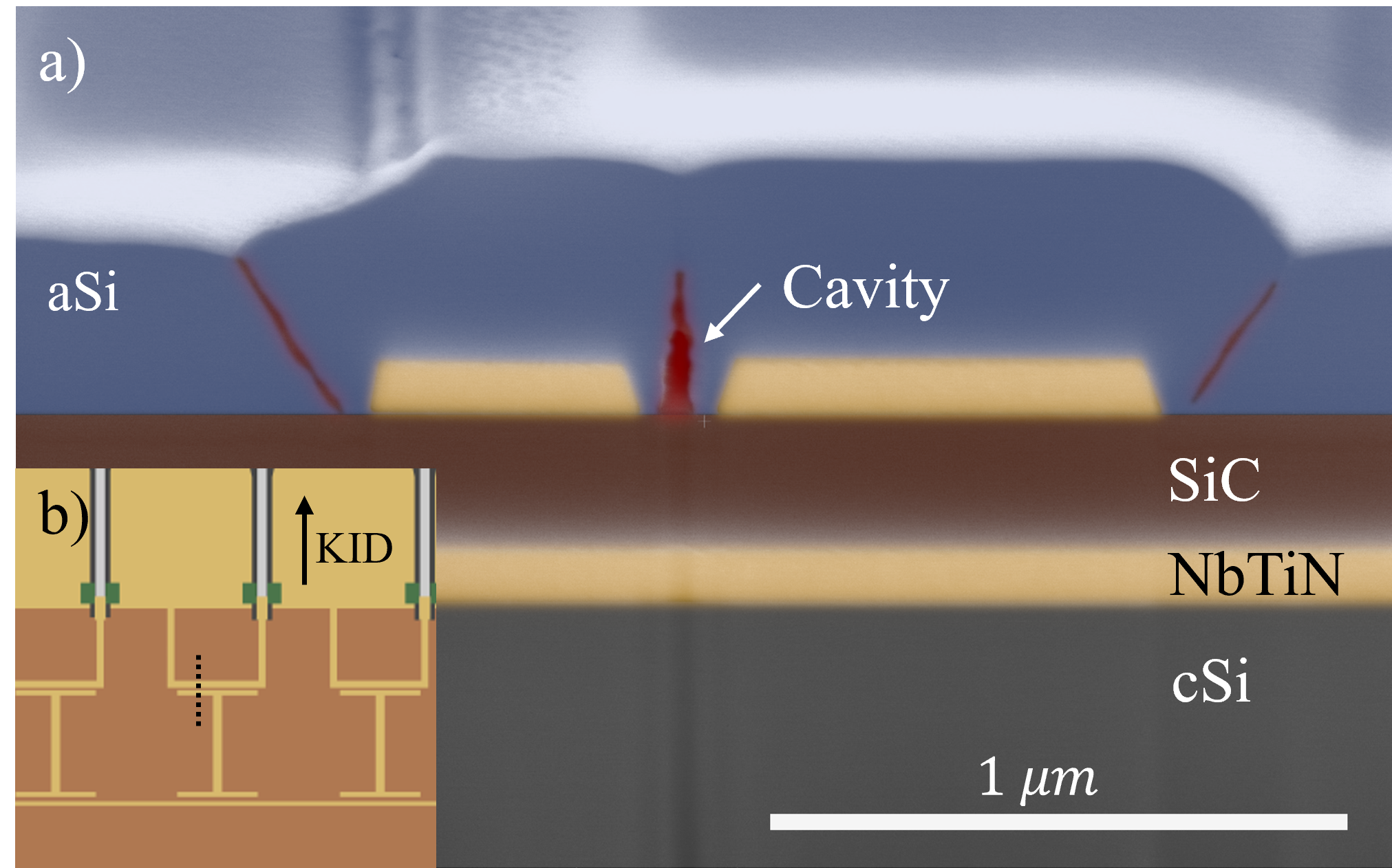}
    \caption{\textbf{a.} the PECVD aSi:H coverlayer (blue) on top of a NbTiN microstrip filter and coupler (yellow) can be used to increase $\epsilon_{eff}$. As can be seen in this FIB image, cavities (red) have formed between the filter and the coupler due to the semi-isotropic growth of the film. \textbf{b.} shows a top view of the structure. The dashed line represents the location of the cross-section. The top of this line corresponds to the right side of the a. For clarity, the cover layer is not shown. }
    \label{fig:filter_cavity}
\end{figure}

We observed a average loaded quality factor ($Q_l$) of 19.4, 25\% lower than the designed value of 25 which was calculated using a 2D approximation of the NbTiN layer. The observed difference can largely be explained by taking into account the coupling of the sidewalls and the dielectric material deposited in the coupling gap. (Marting et al. in prep) Inspection of the coupler gap with a Focused Ion Beam (FIB) and SEM revealed small cavities in between the two sides of the filters, as seen in figure \ref{fig:filter_cavity}. The cavity is thought to be created by the semi-isotropic growth of the PECVD film during the deposition. These cavities partially mitigate the $Q_l$-lowering effect of the capping layer and their rough and inconsistent shape could be a source of Q-scatter. This effect requires future investigation.
Depositions with more anisotropic growth could be explored to minimize the size of the cavity and increase precision in the Q of the microstrip filters. Reduction of the cavity-size can be explored by considering other dielectrics such as SiC\cite{buijtendorpHydrogenatedAmorphousSilicon2022}, by optimizing deposition conditions in the PECVD system or by exploring more directional deposition techniques such as ICPECVD\cite{mungekarHighDensityPlasma2006}.

\section{Microscope exposure for high yield multispaxel-ifu readout lines}
One of the challenges in upscaling a single-spaxel ISS to a many-spaxel IFU capable of spectral mapping is the very long microwave readout line that needs to be routed across multiple spaxels. With IFUs moving towards the wafer scale, this means unperturbed lines up to one meter long. These extended structures will increase the chance of critical defects affecting large sections of the array. This is because a single short along the line prevents the read-out of an entire section of the detector array.

Rather than reducing the particle count of the fabrication process and thereby the number of defects (A process that would require significant investments in the production facility or large changes in the fabrication process), we present a flexible and reliable way of repairing a line short between the ground and the readout line. Comparable to the maskless lithography method described by Gonski and Melngailis \cite{gonskiPhotolithographyUsingOptical2007}.

\begin{figure}[htbp]
    \centering
    \includegraphics[width=\linewidth]{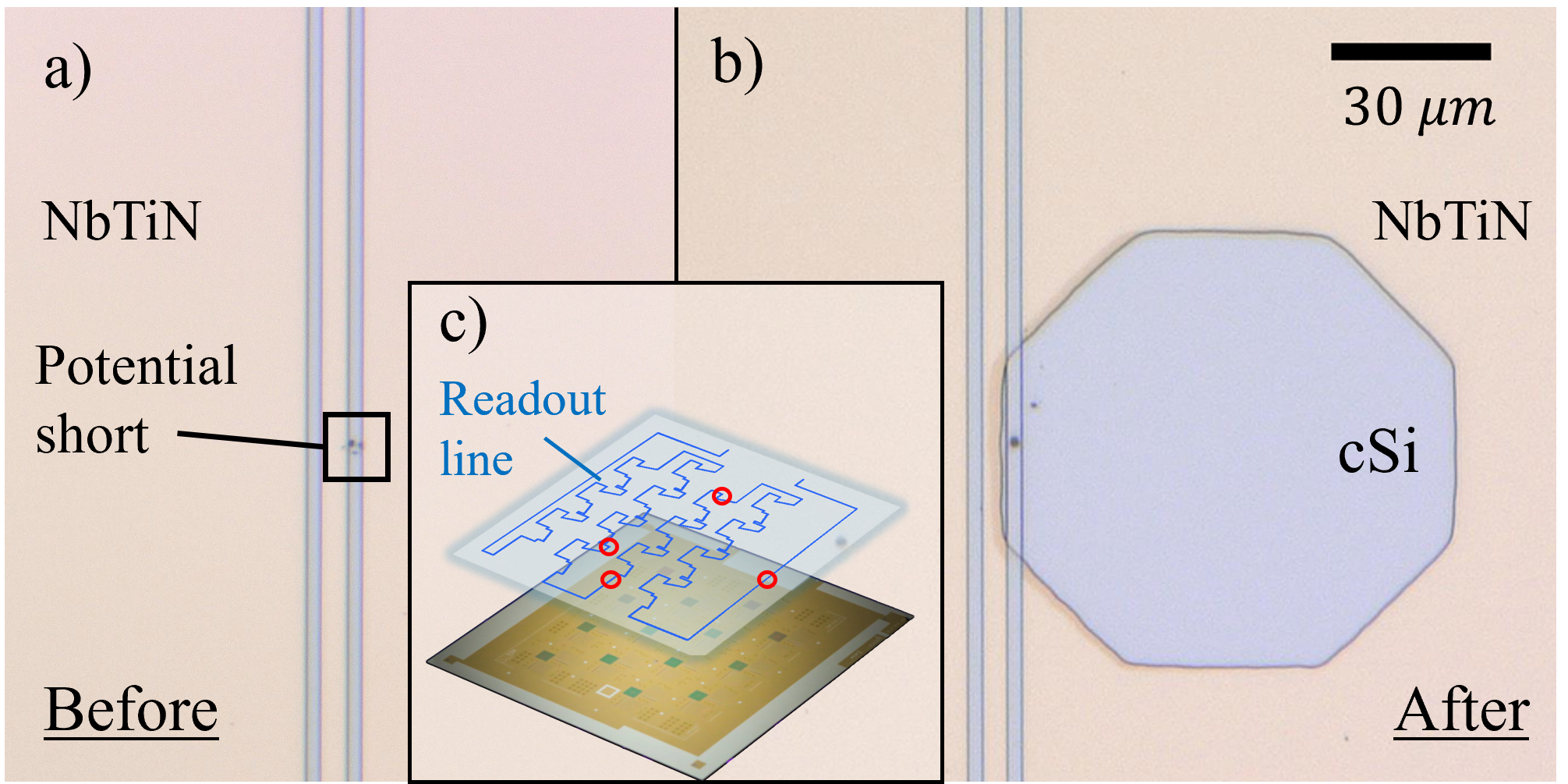}
    \caption{Spaxels on a multispaxel-IFU share long readout lines that, if shorted, will render the detector useless. 
    a) shows a potentially shorted CPW line. b) shows the same area with part of the NbTiN ground plane etched away using microscope exposure. Removing this material isolates the defect and prevents shorting of the line. c) 
    emphasizes the length of the readout line (blue) in comparison to an optical image of a successfully fabricated fourteen-spaxel array. The red circles highlight several potential short sites.}
    \label{fig:linefix}
\end{figure}

After identifying all the defects around the readout line using an optical microscope, we spun a 2.7~$\mathrm{\mu m}$ layer of AZ ECI 3027 photoresist on the substrate. Using a microscope with a blue light filter, we centered the microscope view around the defect. We reduced the aperture of the microscope as much as possible, which left a small opening for the remaining light to travel through. The 180~$\mathrm{\mu m}$-across octagonal spot was then positioned such that it was right against the central line of the CPW, as shown in figure \ref{fig:linefix}.

The optical filter was removed, and the light was set to full illumination. The substrate was exposed for 120 seconds. After dimming the light and putting back the blue light filter, this process was repeated for every short along the CPW line. After the exposure, the resist was developed using AZ 351B, and the exposed area was etched using an SF6/O2 plasma in a Reactive Ion Etcher (RIE). After etching away the conductive material around the defect we stripped away the remaining resist using Acetone. This process restored the shortened signal line to a functioning state, making readout of the array possible again. This fix has resulted in the successful fabrication of a fourteen-spaxel spectrometer array while keeping the wafer intact (Karatsu et al., in prep). Sonnet simulations representing the CPW after material removal, similar to figure \ref{fig:linefix}b), show negligible reflections on isolated CPW-line. Future measurements are planned to confirm this.

\section{Conclusion} 
The microfabrication technology of integrated superconducting spectrometers, such as DESHIMA 2.0, requires several advancements before it is ready to do sensitive and large spectroscopic and polarimetric surveys of the millimeter submillimeter sky. Here, we have presented four such key advances. 
1) An impedance matched aluminum-polyimide microstrip cross-over structure connects each side of a dual polarized leaky wave antenna to their own filterbank. 2) By locally reducing the writing dose by 45\% we prevented the overexposure of a CPW-line across the edge of a SiN membrane. 3) Using a capping layer of dielectric material on top of the filter structures allowed us to lower the quality factor of the filter, however the effect of cavities in the deposited material on the Q-factor of the filters requires further investigation. 4) With a maskless lithography technique employing an optical microscope, we were able to isolate and remove line shorts from a the CPW readoutline to revive a shorted chip.
These advances pave the path for the realization of an on-chip IFU aimed at studying the Cosmic Microwave Background. Using these insights, we have successfully fabricated a fourteen-spaxel IFU (Karatsu et al. in prep).

\section*{Acknowledgment}
We thank the staff of SRON, the Else Kooi Laboratory, and the Kavli Nanolab Delft for their support. This work was supported by the European Union (ERC Consolidator Grant No. 101043486 TIFUUN). Views and opinions expressed are however, those of the authors only and do not necessarily reflect those of the European Union or the European Research Council Executive Agency. Neither the European Union nor the granting authority can be held responsible for them.

\bibliographystyle{IEEEtran}
\bibliography{LTD_TASv4} 

\end{document}